\documentclass[conference, 10pt]{IEEEtran}
\IEEEoverridecommandlockouts

\usepackage[ruled,vlined]{algorithm2e}
\usepackage{amsmath,latexsym,amsfonts,amsthm}
\usepackage{graphicx}
\usepackage{mathtools}
\usepackage{enumitem}
\usepackage{multicol}
\usepackage{lineno,hyperref}
\usepackage{color,colortbl}
\usepackage{float}
\usepackage{subcaption}
\usepackage{balance}
\usepackage{siunitx}
\usepackage{adjustbox}
\usepackage{pifont}
\usepackage{arydshln}
\usepackage{enumitem}
\usepackage{booktabs, multirow} 
\usepackage{soul}

\usepackage{physics}
\usepackage{tikz}
\usepackage{mathdots}
\usepackage{yhmath}
\usepackage{cancel}
\usepackage{color}
\usepackage{array}
\usepackage{gensymb}
\usepackage{tabularx}
\usepackage{extarrows}
\usetikzlibrary{fadings}
\usetikzlibrary{patterns}
\usetikzlibrary{shadows.blur}
\usetikzlibrary{shapes}

\SetKwInOut{Input}{Input}
\SetKwInOut{Output}{Output}
\SetKwInOut{Data}{Data}

\newcolumntype{L}{>{\centering\arraybackslash}m{0.9cm}}
\newcolumntype{M}{>{\centering\arraybackslash}m{2.1cm}}
\newcolumntype{H}{>{\centering\arraybackslash}m{3.0cm}}
\newcolumntype{F}{>{\centering}p{0.18\columnwidth}}
\newcolumntype{A}{>{\centering}p{0.11\columnwidth}}
\newcolumntype{P}{>{\centering}p{0.09\columnwidth}}
\newcolumntype{B}{>{\centering}p{0.07\columnwidth}}

\newcommand{\acronym}{{CELF}}%

\usepackage{cite}
\usepackage{algorithmic}
\usepackage{graphicx}
\usepackage{textcomp}
\usepackage{xcolor}
\def\BibTeX{{\rm B\kern-.05em{\sc i\kern-.025em b}\kern-.08em
    T\kern-.1667em\lower.7ex\hbox{E}\kern-.125emX}}
\begin{document}

\title{Channel Estimation via Loss Field: Accurate Site-Trained Modeling for Shadowing Prediction}



\author{Jie Wang\IEEEauthorrefmark{1}, Meles G. Weldegebriel, and Neal Patwari\\
McKelvey School of Engineering \\
Washington University in St. Louis, Missouri, USA\\
\IEEEauthorrefmark{1}Email: jie.w@wustl.edu}

\maketitle
\pagestyle{plain}

\begin{abstract}

Future mobile ad hoc networks will share spectrum between many users. Channels will be assigned on the fly to guarantee signal and interference power requirements for requested links. Channel losses must be re-estimated between many pairs of users as they move and as environmental conditions change.  Computational complexity must be low, precluding the use of some accurate but computationally intensive site-specific channel models. Channel model errors must be low, precluding the use of standard statistical channel models. We propose a new channel model, \acronym, which uses channel loss measurements from a deployed network in the area and a Bayesian linear regression method to estimate a site-specific loss field for the area.   The loss field is explainable as the site's `shadowing' of the radio propagation across the area of interest, but it requires no site-specific terrain or building information.  Then, for any arbitrary pair of transmitter and receiver positions, \acronym{} sums the loss field near the link line to estimate its channel loss.  We use extensive measurements to show that \acronym{} lowers the variance of channel estimates by up to 56\%. It outperforms 3 popular machine learning methods in variance reduction and training efficiency.
\end{abstract}


\section{Introduction}

Meeting the growing demand for radio spectrum has driven a surge of research to make spectrum allocation increasingly dynamic and shareable \cite{ahmad20205g, bhattarai2016overview}. Examples include spectrum access systems (SAS) which administer the three-tier shared use of the citizens broadband radio service (CBRS) band \cite{sohul2015spectrum}, and the radio dynamic zone \cite{kidd2018national}. 

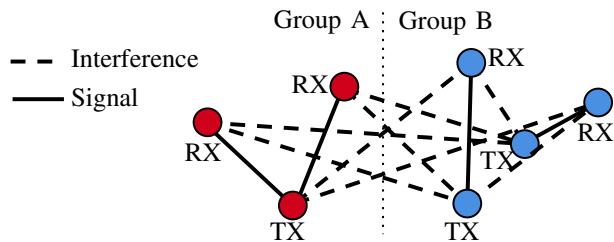
\begin{figure}
    \centering
    \tikzset{every picture/.style={line width=0.75pt}} 

\begin{tikzpicture}[x=0.75pt,y=0.75pt,yscale=-1,xscale=1]

\draw [line width=1.5]    (213,224) -- (237,164) ;
\draw [line width=1.5]    (167,181) -- (213,224) ;
\draw [line width=1.5]    (302,151) -- (300,222) ;
\draw [line width=1.5]    (366,171) -- (324.89,193.68) ;
\draw [line width=1.5]  [dash pattern={on 5.63pt off 4.5pt}]  (167,181) -- (300,222) ;
\draw [line width=1.5]  [dash pattern={on 5.63pt off 4.5pt}]  (238,164) -- (300,222) ;
\draw [line width=1.5]  [dash pattern={on 5.63pt off 4.5pt}]  (238,164) -- (329,192) ;
\draw [line width=1.5]  [dash pattern={on 5.63pt off 4.5pt}]  (167,181) -- (329,192) ;
\draw [line width=1.5]  [dash pattern={on 5.63pt off 4.5pt}]  (213,222) -- (302,151) ;
\draw [line width=1.5]  [dash pattern={on 5.63pt off 4.5pt}]  (213,222) -- (366,171) ;
\draw  [fill={rgb, 255:red, 208; green, 2; blue, 27 }  ,fill opacity=1 ] (231,163) .. controls (231,159.13) and (234.13,156) .. (238,156) .. controls (241.87,156) and (245,159.13) .. (245,163) .. controls (245,166.87) and (241.87,170) .. (238,170) .. controls (234.13,170) and (231,166.87) .. (231,163) -- cycle ;
\draw  [fill={rgb, 255:red, 208; green, 2; blue, 27 }  ,fill opacity=1 ] (205,223) .. controls (205,219.13) and (208.13,216) .. (212,216) .. controls (215.87,216) and (219,219.13) .. (219,223) .. controls (219,226.87) and (215.87,230) .. (212,230) .. controls (208.13,230) and (205,226.87) .. (205,223) -- cycle ;
\draw  [fill={rgb, 255:red, 208; green, 2; blue, 27 }  ,fill opacity=1 ] (162,181) .. controls (162,177.13) and (165.13,174) .. (169,174) .. controls (172.87,174) and (176,177.13) .. (176,181) .. controls (176,184.87) and (172.87,188) .. (169,188) .. controls (165.13,188) and (162,184.87) .. (162,181) -- cycle ;
\draw [line width=1.5]  [dash pattern={on 5.63pt off 4.5pt}]  (302,151) -- (329,192) ;
\draw  [fill={rgb, 255:red, 74; green, 144; blue, 226 }  ,fill opacity=1 ] (322,192) .. controls (322,188.13) and (325.13,185) .. (329,185) .. controls (332.87,185) and (336,188.13) .. (336,192) .. controls (336,195.87) and (332.87,199) .. (329,199) .. controls (325.13,199) and (322,195.87) .. (322,192) -- cycle ;
\draw  [fill={rgb, 255:red, 74; green, 144; blue, 226 }  ,fill opacity=1 ] (295,151) .. controls (295,147.13) and (298.13,144) .. (302,144) .. controls (305.87,144) and (309,147.13) .. (309,151) .. controls (309,154.87) and (305.87,158) .. (302,158) .. controls (298.13,158) and (295,154.87) .. (295,151) -- cycle ;
\draw [line width=1.5]  [dash pattern={on 5.63pt off 4.5pt}]  (300,222) -- (366,171) ;
\draw  [fill={rgb, 255:red, 74; green, 144; blue, 226 }  ,fill opacity=1 ] (359,171) .. controls (359,167.13) and (362.13,164) .. (366,164) .. controls (369.87,164) and (373,167.13) .. (373,171) .. controls (373,174.87) and (369.87,178) .. (366,178) .. controls (362.13,178) and (359,174.87) .. (359,171) -- cycle ;
\draw  [fill={rgb, 255:red, 74; green, 144; blue, 226 }  ,fill opacity=1 ] (293,222) .. controls (293,218.13) and (296.13,215) .. (300,215) .. controls (303.87,215) and (307,218.13) .. (307,222) .. controls (307,225.87) and (303.87,229) .. (300,229) .. controls (296.13,229) and (293,225.87) .. (293,222) -- cycle ;
\draw  [dash pattern={on 0.84pt off 2.51pt}]  (257.5,125.5) -- (257.5,237.45) ;
\draw [line width=1.5]  [dash pattern={on 5.63pt off 4.5pt}]  (69.67,150.33) -- (95.56,150.33) ;
\draw [line width=1.5]    (70.67,169.33) -- (96.56,169.33) ;

\draw (210,156) node [anchor=north west][inner sep=0.75pt]   [align=left] {RX};
\draw (309,142) node [anchor=north west][inner sep=0.75pt]   [align=left] {RX};
\draw (287,230) node [anchor=north west][inner sep=0.75pt]   [align=left] {TX};
\draw (199,230) node [anchor=north west][inner sep=0.75pt]   [align=left] {TX};
\draw (156,190) node [anchor=north west][inner sep=0.75pt]   [align=left] {RX};
\draw (305,194) node [anchor=north west][inner sep=0.75pt]   [align=left] {TX};
\draw (354,180) node [anchor=north west][inner sep=0.75pt]   [align=left] {RX};
\draw (201,123) node [anchor=north west][inner sep=0.75pt]   [align=left] {Group A};
\draw (264,123.37) node [anchor=north west][inner sep=0.75pt]   [align=left] {Group B};
\draw (98.67,142.33) node [anchor=north west][inner sep=0.75pt]   [align=left] {Interference};
\draw (98.67,162.33) node [anchor=north west][inner sep=0.75pt]   [align=left] {Signal};

\end{tikzpicture}
    \caption{Assigning channels and transmit powers to ensure required SIRs among links between $T$ transmitters (TX) and $R$ receivers (RX) requires $RT$ channel loss estimates, recomputing each position change for mobile devices.}
    \label{fig:sir_diagram}
\end{figure}

It is a challenge to achieve real time, accurate, and efficient dynamic spectrum allocation; a major part of this challenge is to accurately and continuously predict signal and interference powers on all pairs of proximate mobile devices, as shown in Fig.~\ref{fig:sir_diagram}. When transmit powers are known, knowing these received powers reduces to \textit{channel estimation}, also known as path loss prediction. Current channel models are not well-matched to the needs of dynamic spectrum management in mobile networks.

\begin{figure*}
    \centering
    \includegraphics[width=\textwidth]{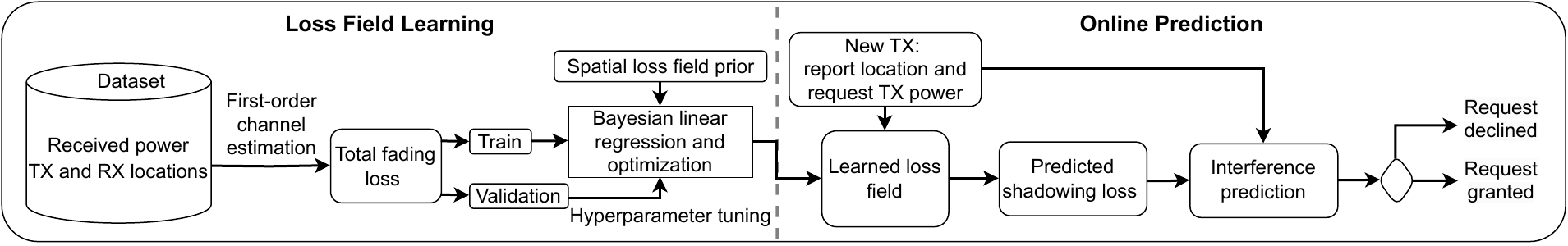}
    \caption{Overview of the proposed \acronym{} model and the online shadowing prediction process.}
    \label{fig:lossfieldimaging_diagram}
\end{figure*}



Many current path loss prediction models require computing radio propagation mechanisms such as reflection and diffraction in the particular geometry of the area of the network deployment. Ray tracing models \cite{yun2015ray} need high-resolution building databases and are extremely computationally complex.  The terrain-integrated rough earth model (TIREM)  model \cite{eppink1994tirem} considers a profile of the terrain features and building heights on the line between the transmitter and receiver. However, computing diffraction losses on this profile is also computationally intensive. While such models improve model accuracy compared to general models which curve-fit to empirical data, such as Okumura-Hata  \cite{hata1980empirical} and log-distance path loss exponent \cite{rappaport2010wireless} models, their computational requirements significantly limit the adaptability of today's spectrum management systems.  Emerging machine learning (ML) channel models can be both accurate and fast during testing, but require very large datasets and high computational complexity during model training. ML models offer no explanation or reasoning for its predictions, which prevents system engineers from diagnosing problems when the model performs poorly. Finally, general-purpose ML models are largely unable to accurately account for the underlying physical mechanisms which cause link measurements to be correlated in space.

In this paper, our contribution is to develop a new type of channel learning model, the Channel Estimation via Loss Field (\acronym{}) model, which achieves the best of multiple types of channel models.  
\acronym{} formulates the link fading loss as a linear function of a shadowing loss field. 
As shown in Fig.~\ref{fig:lossfieldimaging_diagram}, this loss field accounts for the physical mechanism of shadowing due to obstacles in the spatial domain. The loss field is learned from training measurements via Bayesian linear regression, but training is lower in computation requirements compared to a general-purpose ML model. Using training measurements allows the model to fit the particular site of deployment.  Sensors deployed as part of a radio dynamic zone, or conducted by nodes as part of the dynamic spectrum access protocol, can be used for this purpose; and training data quantities can be low in comparison with other ML methods.  We consider the stability and optimization of the \acronym{} algorithm to achieve a robust algorithm with a low computational cost. To predict shadowing loss for a new link, the model only needs to compute a weighted sum of the learned loss image in which most weights are zero.

We use outdoor and indoor datasets to experimentally validate how accurately and efficiently \acronym{} can perform. We compare \acronym{} with three general-purpose ML methods: SVR, random forests, and multi-layer perceptron (MLP)-ANN, in terms of (1) variance reduction, (2) training efficiency, and (3) prediction efficiency. \acronym{} reduces the variance of total fading loss by up to 56\% outdoors and 40\% indoors. In comparison to the ML-based methods, \acronym{} achieves larger variance reductions. The MLP-ANN model is the most accurate model out of the three ML-based methods, but it requires three times more time than \acronym{} for model training. For shadowing loss prediction, test dataset size and the loss field size severely impact \acronym{}'s prediction efficiency. It requires on average 54.8 times more prediction time than MLP-ANN for our large outdoor dataset, and is 2.5 times slower than MLP-ANN for the indoor dataset.

For perspective, channel models \textit{do not predict small-scale fading effects}, i.e., those caused by sub-wavelength (cm-level) changes in the position of the transmitter or receiver.  Small-scale fading is severe (e.g., more than 20 dB 1\% of the time in a Rayleigh fading channel \cite{rappaport2010wireless}) and not predicted by channel models, which don't know device and environmental obstruction positions to that level of accuracy. Instead, channel models like our proposed model predict large-scale fading (caused by increasing distance) and medium-scale or shadow fading (caused by obstructions).  Since our measurements include small-scale fading but our model cannot predict it, we cannot reduce the estimate variance to zero.  Instead, we judge models by how much they can reduce fading variance compared to a standard statistical channel model, and we find that \acronym{} does better than any other learning-based model.

\section{Channel Estimation via Loss Field}

In this section, we first present a base model for the first-order estimate of the channel. It is followed by spatial loss field learning which characterizes shadowing loss for improved channel estimation. Finally, we discuss small-scale fading and noise which are commonly random and unpredictable.

\subsection{First-order channel estimation}\label{sec: pathlossfading}
The base model for the first-order channel estimation can be arbitrary. It could be any model listed in the related work (Section~\ref{sec:related}). In this work, we choose the \textit{log-distance path loss} model as the base model due to its 
computational simplicity and universality of frequency and environment.

The \textit{log-distance path loss} model \cite{rappaport2010wireless, goldsmith2005wireless} states that the ensemble average power $\Bar{P}(d_{l})$ along a link $l=(i, j)$ between node $i$ and node $j$ reduces in a logarithmic manner as the distance increases, 
\begin{equation}\label{model: pathloss} 
    \begin{aligned}
        & \Bar{P}(d_{l}) = P_T - \Pi_0 - 10n_p\log\frac{d_{l}}{\Delta_0}
    \end{aligned}
\end{equation}
where $P_T$ is the transmitted power in dBm, $d_{l}$ is the link distance, and $\Pi_0$ is a constant specifying the dB loss at a reference distance $\Delta_0$. $n_p$ is the path loss exponent which indicates the level of environmental clutter.

Given the same distance $d_{l}$, the received power measurements vary around the average $\Bar{P}(d_{l})$ due to shadow fading and small-scale fading \cite{goldsmith2005wireless}. As a result, the received power $P(d_{l})$ along the link $l$ can be written as:
\begin{equation}\label{model: fading} 
    \begin{aligned}
        P(d_{l}) &= \Bar{P}(d_{l}) - Z_{l}\\
        Z_{l} &= X_{l} + Y_{l}
    \end{aligned}
\end{equation}
where $Z_{l}$ is the total fading loss. It consists of independent shadowing loss $X_{l}$ and small-scale fading loss $Y_{l}$ \cite{wilson2010radio}. 

\subsection{Network shadowing model for shadowing correlation}

The total fading loss $Z_{l}$ is commonly modeled as independent and identically distributed ($i.i.d.$) across links \cite{bettstetter2003connectivity,hekmat2006connectivity,chen2011implications}. However, it is in contrast to the empirical observation that shadowing losses along two links are correlated due to obstructions, e.g., outdoor buildings and terrain variations, and indoor walls and furniture \cite{gudmundson1991correlation, Piyush2009Correlated, lee2018effect}. 

In order to simultaneously model the correlations in shadow fading that exist across multiple link pairs in a network, we use the network shadowing model of \cite{patwari2008effects}. Let $\mathcal{L}$ be a set of link pairs in a wireless network, and $L=|\mathcal{L}|$ is the size of the set. We assume that each link is different in either transmitter or receiver from the other links in the set $\mathcal{L}$. The network shadowing model describes the joint link fading loss as:

\begin{equation}\label{totalfading: matrixform} 
    \boldsymbol{z} = \boldsymbol{W}\boldsymbol{p}+\boldsymbol{n}
\end{equation}
 where $\boldsymbol{z} = [Z_{1}, Z_{2}, \ldots, Z_{L}] \in \mathcal{R}^{L\times 1}$ is the total fading loss vector, $\boldsymbol{W} \in \mathcal{R}^{L \times M}$ is a weight matrix, $\boldsymbol{p} \in \mathcal{R}^{M \times 1}$ is a discretized loss field in dB, and 
$\boldsymbol{n} \in \mathcal{R}^{L\times 1}$ is a measurement noise vector. Their details are given below. 

\textbf{Spatial loss field}. The spatial loss field is introduced in \cite{patwari2008effects}. it characterizes the environment of interest as a Gaussian random field, represented as a vector $\mathbf{p}$ with $M$ pixels, which is isotropic wide-sense stationary. It has zero mean and the following exponentially decaying spatial covariance function: 

\begin{equation}\label{shadowing: lossfield} 
    \begin{aligned}
        C_p(m,n) = \frac{\sigma_X^2}{\delta} \exp (-\frac{d_{m,n}}{\delta})
    \end{aligned}
\end{equation}
where $d_{m,n}$ is the Euclidean distance between the centers of pixels $m$ and $n$, $\sigma^2_X$ is the variance of the shadowing loss, and $\delta$ is a space constant to be determined.

The shadowing loss on a link $l$ then is a weighted sum of the loss field $\boldsymbol{p}$ over the pixels that cross the link $l$.



\begin{figure}
    \centering
    \includegraphics[width=0.6\columnwidth]{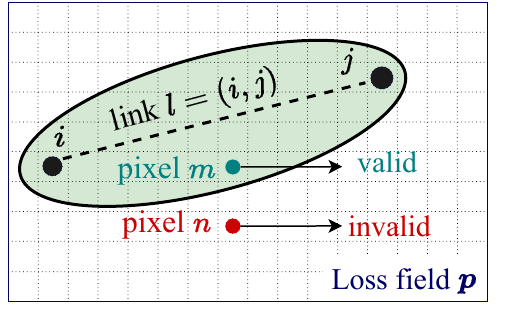}
    \caption{The ellipse model for selecting valid pixels ({\color{teal} $\bullet$}) that contribute to the shadowing loss of link $l=(i, j)$.}
    \label{fig: weight_model}
\end{figure}

\textbf{Weight matrix model}.\label{weightmodel}
A popular ellipse model in \cite{wilson2010radio} is adopted for the weight matrix $\boldsymbol{W}$, as shown in Fig.~\ref{fig: weight_model}. It considers the two ends of link $l$ as the foci and utilizes a tunable parameter $\lambda$ to determine the ellipse width. A pixel is viewed as valid if it falls within the ellipse, and the corresponding weight in $\boldsymbol{W}$ will have a non-zero contribution to the shadowing loss of link $l$. Past studies \cite{patwari2008effects, Piyush2009Correlated, wilson2010radio}, based on propagation physics, have constructed the weight to be:
\begin{equation}\label{totalfading:weightmatrix} 
    w_{lm} = \frac{1}{\sqrt{d_l}}
    \begin{cases}
        1, & \mbox{if } d_{l,m}{(1)} + d_{l,m}{(2)} < d_l +\lambda\\
        0, &\mbox{otherwise}
    \end{cases}
\end{equation}
where $d_{l,m}{(1)}$ and $d_{l,m}{(2)}$ are the distances from the center of pixel $m$ to the two ends of link $l$, $d_l$ is the link distance, and $\lambda$ denotes the width of the ellipse.



\textbf{Small-scale fading and noise}. The Gaussian noise $\boldsymbol{n}$ includes both small-scale fading loss and measurement noise which are independent of each other. Small-scale fading is also known as multipath fading which describes the attenuation that occurs from constructive and destructive addition of multipath signals \cite{goldsmith2005wireless}. While the multipath channel is commonly modeled as Rayleigh and Rician distributions \cite{tse2005fundamentals}, we follow the rationale in \cite{Piyush2009Correlated} and characterize it as $i.i.d.$ Gaussian in dB.

\subsection{Loss field learning}\label{sec: lossfieldimaging}
\textbf{Bayesian linear regression}.
Given the linear joint link model in (\ref{totalfading: matrixform}) and the Gaussian loss field prior, we can reconstruct the loss field $\boldsymbol{p}$ via Bayesian linear regression. To start with, the likelihood function of the total fading loss vector is as follows:
\begin{equation}\label{likelihood}
    \begin{aligned}
        f(\boldsymbol{z}|\boldsymbol{W}, \boldsymbol{p}, \sigma^2_n) = \mathcal{N}(\boldsymbol{W}\boldsymbol{p}, \sigma^2_n \boldsymbol{I}_L)
    \end{aligned}
\end{equation}
Next, the loss field prior is modeled as Gaussian with its probability density function (pdf) as:
\begin{equation}\label{prior}
    \begin{aligned}
        f(\boldsymbol{p}) = \mathcal{N}(\boldsymbol{0}, \boldsymbol{C_p})
    \end{aligned}
\end{equation}
Therefore we can learn that the posterior pdf of $\boldsymbol{p}$ is multivariate Gaussian as 
\begin{equation}\label{posterior}
    \begin{aligned}
        f(\boldsymbol{p}|\boldsymbol{z}, \boldsymbol{W}, \sigma^2_n) &\propto  f(\boldsymbol{z}|\boldsymbol{W}, \boldsymbol{p}, \sigma^2_n) \cdot f(\boldsymbol{p})\\
        &= \mathcal{N}(\boldsymbol{\mu}_{\boldsymbol{p}|\boldsymbol{z}}, \boldsymbol{C}_{\boldsymbol{p}|\boldsymbol{z}})
    \end{aligned}
\end{equation}
where
\begin{equation}\label{posterior:mean_and_cov}
    \begin{aligned}
        &\boldsymbol{\mu}_{\boldsymbol{p}|\boldsymbol{z}} = \sigma^{-2}_n (\sigma^{-2}_n\boldsymbol{W}^T \boldsymbol{W} + \boldsymbol{C}^{-1}_{\boldsymbol{p}})^{-1} \boldsymbol{W}^T \boldsymbol{z},\\ 
        &\boldsymbol{C}_{\boldsymbol{p}|\boldsymbol{z}} = (\sigma^{-2}_n\boldsymbol{W}^T\boldsymbol{W} + \boldsymbol{C}^{-1}_{\boldsymbol{p}})^{-1}
    \end{aligned}
\end{equation}
As a result, we can acquire both the minimum mean-square error (MMSE) solution and the maximum a posterior (MAP) estimator $ \hat{\boldsymbol{p}}$ as the posterior mean $\boldsymbol{\mu}_{\boldsymbol{p}|\boldsymbol{z}}$ in (\ref{posterior:mean_and_cov}).

The solution can be further generalized by a tunable regularizer $\alpha$, which leads to the final estimator $\hat{\boldsymbol{p}}$ as:
\begin{equation}\label{image_estimator}
    \begin{aligned}
         \hat{\boldsymbol{p}} &= \Pi_1 \boldsymbol{z}\\
         \Pi_1 &= (\boldsymbol{W}^T \boldsymbol{W} + \alpha \boldsymbol{C}^{-1}_{\boldsymbol{p}})^{-1} \boldsymbol{W}^T 
    \end{aligned}
\end{equation}

\textbf{Solution stability}.
The linear regression, however, is an ill-posed problem, i.e., the attenuation image estimate $\hat{\boldsymbol{p}}$ from the measurement vector is not unique. Such ill-posedness is due to two main factors: \\
(1) $L<M$: there are more pixels to be estimated than link measurements, which makes the problem underdetermined;\\
(2) $L>M$ but with a sparse $\boldsymbol{W}$: only a few pixels are assigned non-zero weights for each link and thus $\boldsymbol{W}$ is still rank-deficient regardless of the number of link samples.

For a stable solution, the regularization constant $\alpha$ in (\ref{image_estimator}) is required to be positive. In doing so, the estimator is robust to rank deficiency in the weight matrix, and the inverse term in the operator $\Pi_1$ always exists.

\textbf{Solution efficiency}.
Latency can be the other concern given large datasets and wide area estimation, and thus requires efficiency improvement. If $L<M$, we can review the problem as sparse linear regression and adopt the common minimum norm (MNE) solution as:
\begin{equation}\label{minimum_norm}
    \begin{aligned}
        \hat{\boldsymbol{p}} &= \Pi_2 \boldsymbol{z}\\
         \Pi_2 &= \boldsymbol{C}_{\boldsymbol{p}} \boldsymbol{W}^T (\boldsymbol{W}\boldsymbol{C}_{\boldsymbol{p}}\boldsymbol{W}^T+\alpha \boldsymbol{I})^{-1}
    \end{aligned}
\end{equation}
which calculates an inverse of only a $\mathcal{R}^{L \times L}$ matrix rather than $\mathcal{R}^{M \times M}$.

If $L>M$, we leverage the Cholesky decomposition \cite{golub2013matrix} to lower the latency. It is based on the fact that $(\boldsymbol{W}^T \boldsymbol{W} + \alpha \boldsymbol{C}^{-1}_{\boldsymbol{p}})$ in $\Pi_1$ is symmetric and positive definite. Let $\boldsymbol{A} = \boldsymbol{W}^T \boldsymbol{W} + \alpha \boldsymbol{C}^{-1}_{\boldsymbol{p}}$, and $\boldsymbol{b} = \boldsymbol{W}^T \boldsymbol{z}$. We first calculate the triangular matrix $\boldsymbol{S}$ via the Cholesky factorization:
\begin{equation}\label{Cholesky_fact}
    \begin{aligned}
        \boldsymbol{S}\boldsymbol{S}^T = \boldsymbol{A}, \quad \boldsymbol{S} = \text{chol} \boldsymbol{A}.
    \end{aligned}
\end{equation}
By reformulating the problem as $\boldsymbol{S}\boldsymbol{S}^T\boldsymbol{p} = \boldsymbol{b}$, the loss field estimate $\hat{\boldsymbol{p}}$ can be obtained via forward-backward substitution. According to \cite{garnett2023bayesian},  the Cholesky decomposition can be twice as efficient as the general LU decomposition.

\section{Evaluation Methodology}\label{sec: implementation}

In this section, we describe one outdoor and one indoor real-world received power dataset, three popular ML-based methods, and two evaluation metrics for assessing the proposed \acronym{} algorithm's performance. 


\begin{figure*}
  \begin{subfigure}[b]{0.33\textwidth}
    \includegraphics[width=\columnwidth]{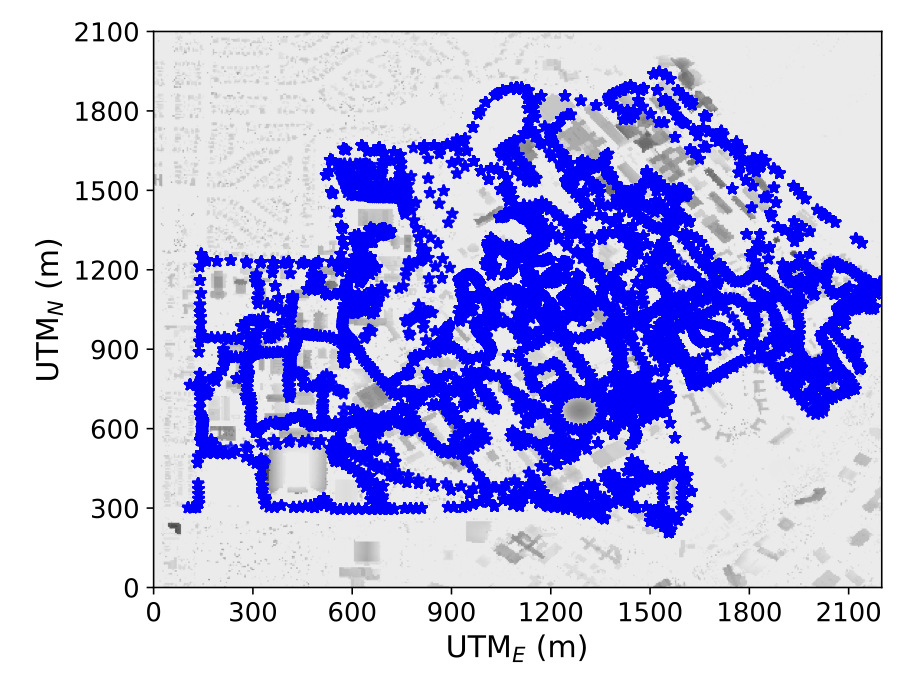}
    \caption{The outdoor transmitter locations}
    \label{fig:frs_dataset_transmitter}
  \end{subfigure}
  \begin{subfigure}[b]{0.33\textwidth}
    \includegraphics[width=\columnwidth]{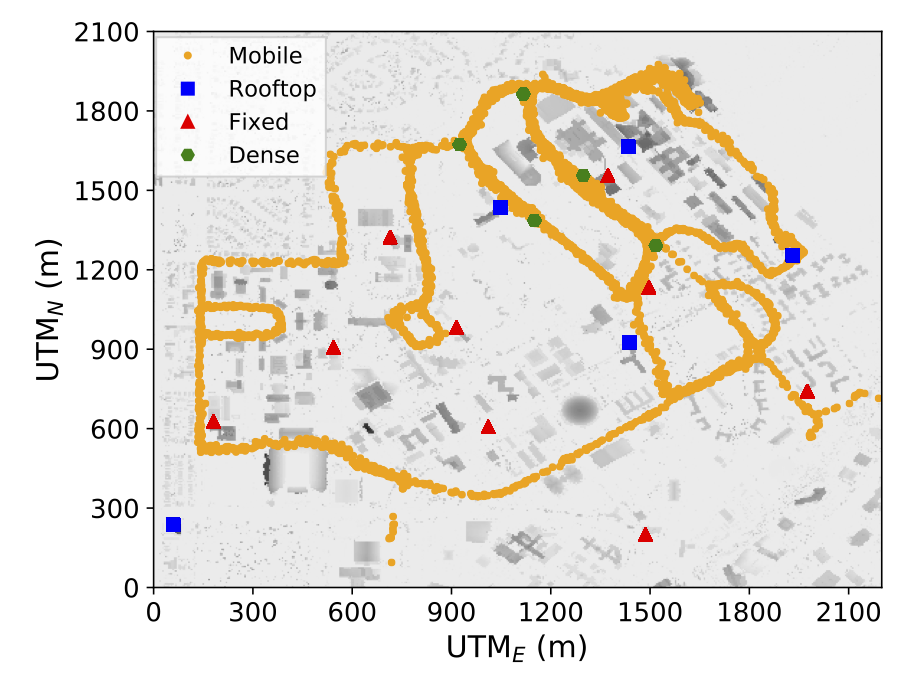}
    \caption{The outdoor receivers' locations}
    \label{fig:frs_dataset_receiver}
  \end{subfigure}
  \begin{subfigure}[b]{0.33\textwidth}
    \includegraphics[width=\columnwidth]{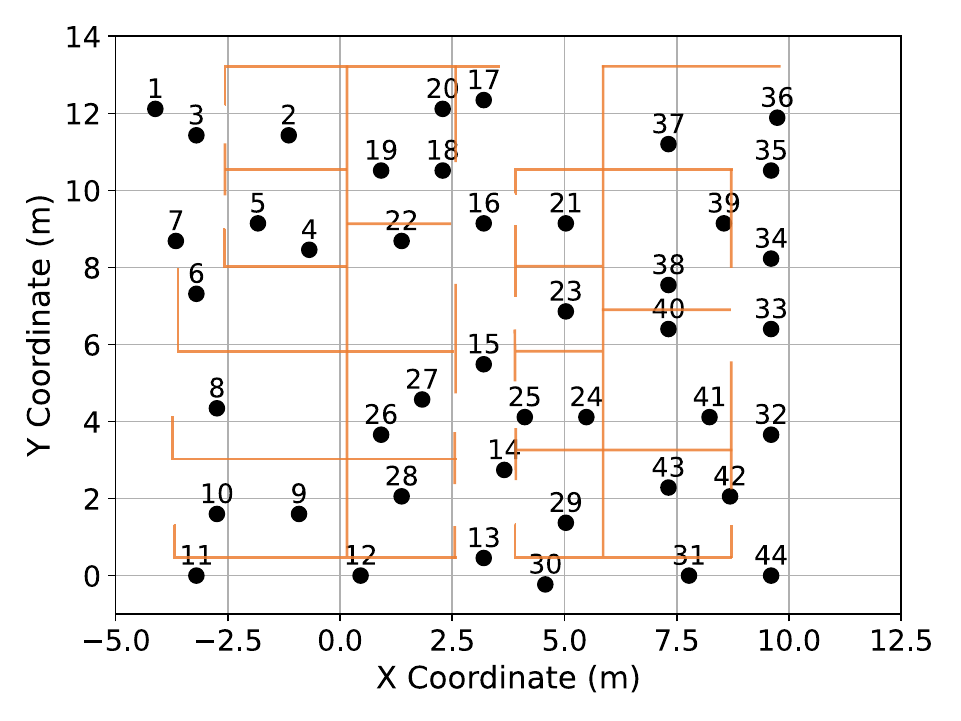}
    \caption{The 44 indoor node locations}
    \label{fig:indoor_dataset}
  \end{subfigure}
\caption{The transmitter and receiver locations of the outdoor and indoor received power datasets. They are collected in a 2200$\times$2100~m$^2$ campus area and a 17.5$\times$15~m$^2$ indoor office with cubicles ({\color{orange} --}) respectively.}
\label{fig: frs_indoor_dataset}
\end{figure*}

\begin{figure*}[t]
  \begin{subfigure}[b]{0.49\textwidth}
    \includegraphics[width=\textwidth]{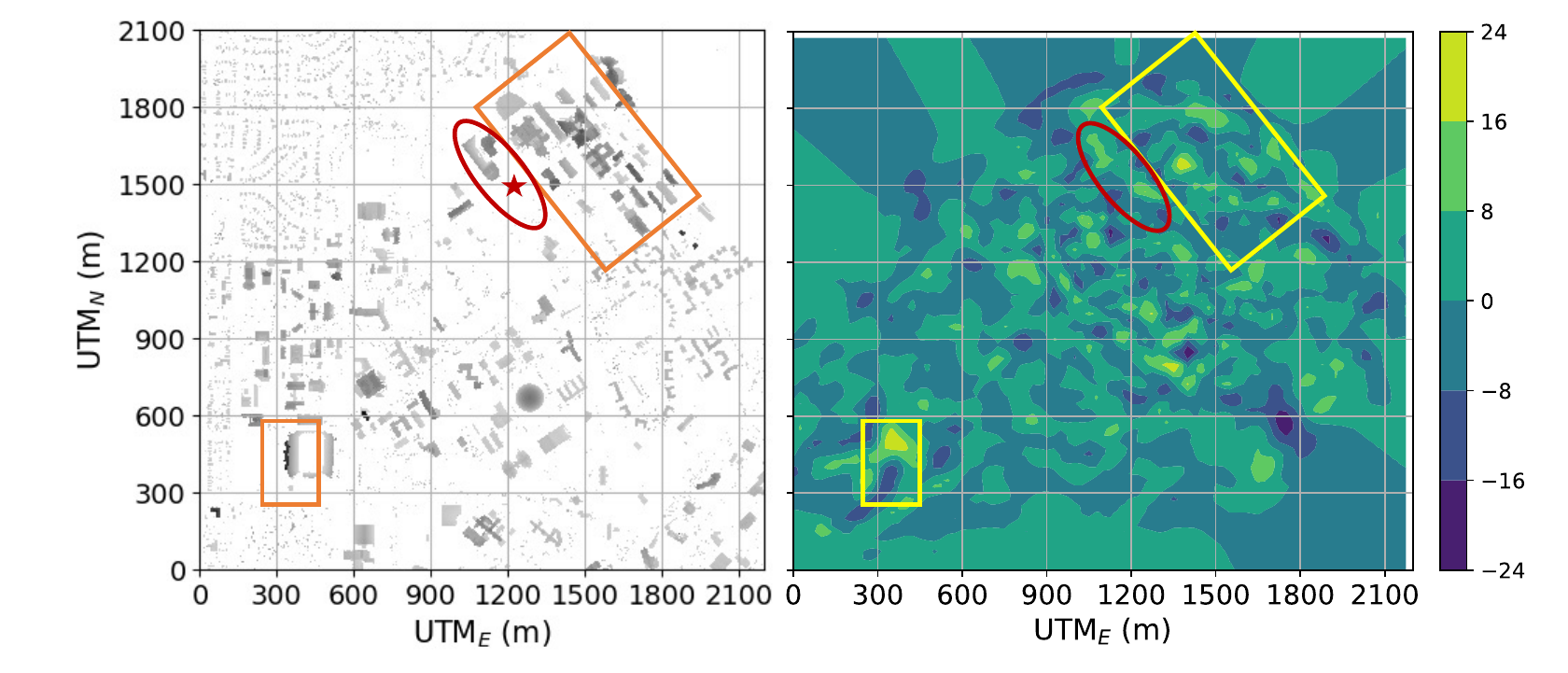}
    \caption{the \textit{Rooftop} training dataset.}
    \label{fig:cbrssdr_LossField}
  \end{subfigure}
  \hfill 
  \begin{subfigure}[b]{0.49\textwidth}
    \includegraphics[width=\textwidth]{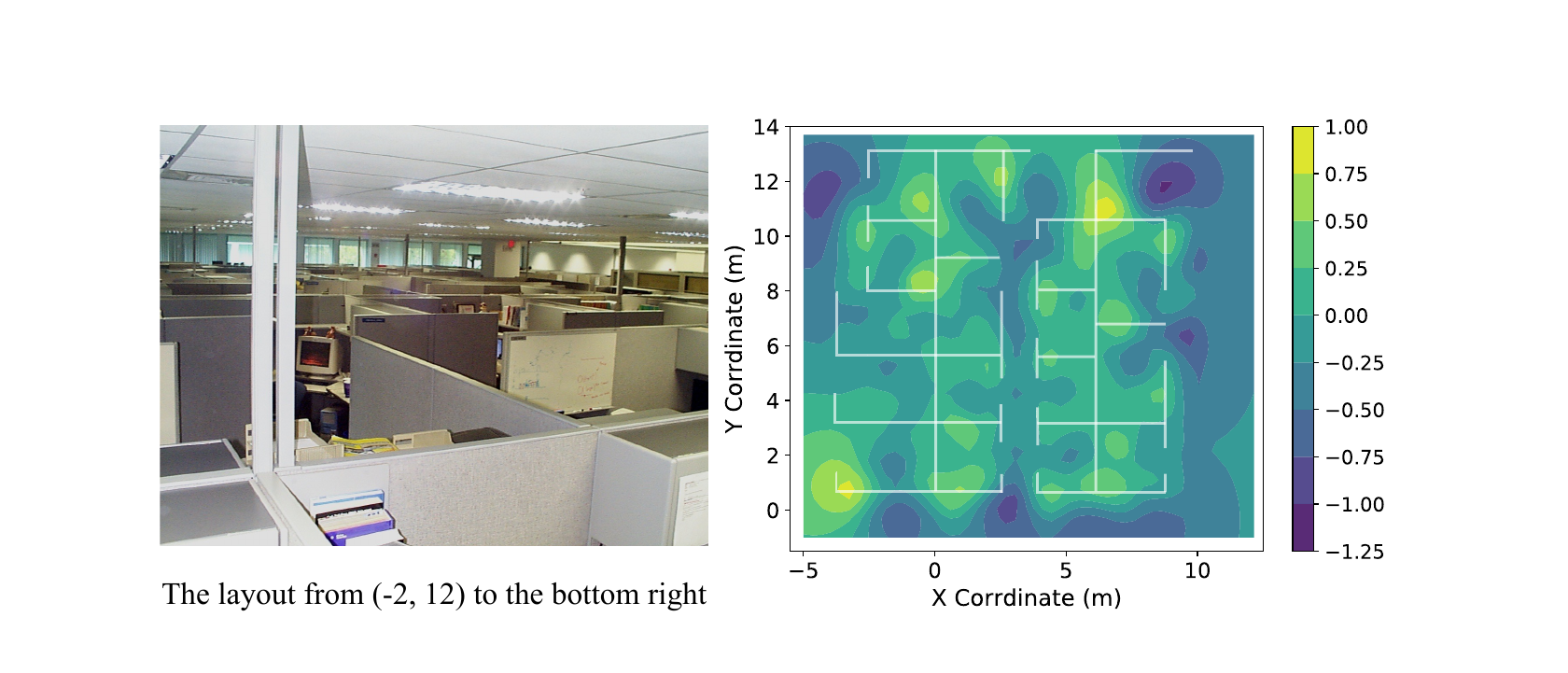}
    \caption{the indoor training dataset.}
    \label{fig:indoor_LossField}
  \end{subfigure}
\caption{Example loss images learned via the proposed \acronym{} algorithm and the site maps as a reference.}
\label{fig: LossField}
\end{figure*}

\subsection{Real-world received power datasets}\label{subsec: datasets}

\textbf{Outdoor Dataset.}
This dataset \cite{RFSData2022gitlab} is from a $2200\times2100 m^2$ university campus area. A portable commercial radio is used as the transmitter, and the receivers are 25 software-defined radio (SDR) nodes with omnidirectional antennas deployed on POWDER, an open wireless experimental testbed \cite{Breenwintech20}. These receivers can be categorized into 4 receiver types, \textit{Rooftop, Fixed, Mobile, and Dense}, according to the radio-antenna-placement differentiation. Table~\ref{results: outdoordataset_spec} gives specifications for each receiver type. Fig.~\ref{fig:frs_dataset_transmitter} and \ref{fig:frs_dataset_receiver} show the GPS coordinates of the transmitter and all the receivers on the campus map. As the four types of receivers are heterogeneous and uncalibrated, this work treats the data collected by each type as a separate dataset. 
\begin{table}
  {\begin{tabular}{AABAFL} 
    \toprule
     Dataset &  Receiver & Count & SDR &
      Antenna Height (m) & Samples\\
    \midrule
    \multirow{ 4}{*}{Outdoor}  &  Rooftop & 5 & X310 &  28--51  & 13114 \\
    & Fixed & 12 & B210 &  1.5 & 24253\\
    & Mobile & 7 & B210 &  2.0 & 8688\\
    & Dense & 5 & B210 &  8.8 & 13268\\
    \midrule
    Indoor & DS-SS & 44 & -- & 1.0 & 946\\
    \bottomrule
  \end{tabular}}
  \caption{Specifications for the two datasets.}
  \label{results: outdoordataset_spec}
\end{table}


\textbf{Indoor Dataset.}
This dataset \cite{indoor2007patwari} is from in an indoor office area, a 17.5~m $\times$ 15~m space surrounded by 1.8~m high cubicle walls. Channels between all pairs of 44 device locations and the cubicle wall positions are shown in Fig.~\ref{fig:indoor_dataset}. This indoor dataset has in total $44\times 43 \times 0.5 = 946$ measurements, which are described in detail in \cite{patwari2003relative}. 

\textbf{Train-Test Split.}\label{impl: split}
 Each dataset needs to be split without overlapping for loss field estimation (training) and shadowing loss prediction (testing) purposes. We choose the link index as the criterion to partition the datasets. Each dataset is split with a 7: 3 ratio. Each data point is randomly assigned for training or testing.

 \subsection{Methods for comparison}\label{subsec: ML_methods}

We adopt the \textit{Okumura-Hata model} and three general-purpose ML models, \textit{Random Forest, SVR, and MLP-ANN}, in this work for performance comparison. The rationale behind such choices is: (1) they represent the two main categories in the related work -- non-learning and learning approaches; (2) they require neither site-specific terrain information nor large-scale datasets, unlike complex deep learning models such as RadioUNet \cite{levie2021radiounet} and PL-GAN \cite{marey2022pl}; (3) they have been widely used as benchmarks for path loss prediction \cite{goldsmith2005wireless, wen2019path, zhang2019path, wu2020artificial}, and the \textit{Okumura-Hata model} particularly has been in use for the CBRS band sharing and analysis \cite{drocella20163}.
\begin{itemize}
    \item \textit{Okumura-Hata} \cite{hata1980empirical}: it provides a closed-form empirical formula for path loss computation over 150-1500 MHz frequency range. This model is only compared across outdoor datasets as it does not capture indoor environments.
    \item \textit{Random Forest} \cite{oroza2017machine}: it is an ensemble learning approach that first constructs multiple decision trees on random subsets of the dataset and then combines them to improve the accuracy and robustness of the model.
    \item \textit{SVR} \cite{moraitis2020machine}: it is a variation of support vector machines used for regression. Unlike traditional squared error minimization, SVR fits a line or a curve by maximizing the margin of error.
    \item \textit{MLP-ANN} \cite{wu2020artificial}: MLP-ANN is a feedforward neural network that consists of an input layer, an output layer, and multiple hidden layers. It is trained iteratively using algorithms like stochastic gradient descent for squared error optimization.
\end{itemize}

 \subsection{Evaluation metrics}\label{model:evaluation}

We adopt two evaluation metrics, \textit{variance reduction} and \textit{running time}, to quantify the performance of the tested algorithms. To specify, \textit{variance reduction} is defined as the percentage decrease of the fading loss variance, i.e., 
\begin{equation}\label{evaluation: accuracy_improvement}
    \begin{aligned}
        \epsilon = \frac{\sigma^2_{{\boldsymbol{z}_{\mathcal{T}}}} - \sigma^2_{{\text{err}}}}{\sigma^2_{{\boldsymbol{z}_{\mathcal{T}}}}}\cdot 100\%
    \end{aligned}
\end{equation}

where $\sigma^2_{{\boldsymbol{z}_{\mathcal{T}}}}$ is the fading loss variance of a dataset ${\mathcal{T}}$ after the first-order channel estimation in Section \ref{sec: pathlossfading}, and $\sigma^2_{{\text{err}}}$ is the error variance after shadowing loss subtraction which is computed as the mean-squared error (MSE):
\begin{equation}\label{evaluation: accuracy_improvementmse}
    \begin{aligned}
        \sigma^2_{{\text{err}}} = \frac{\|{\boldsymbol{z}_{\mathcal{T}}} - \boldsymbol{W}_{\mathcal{T}} \cdot \hat{\boldsymbol{p}}\|^2}{N_{\mathcal{T}}}
    \end{aligned}
\end{equation}

where $\hat{\boldsymbol{p}}$ is the attenuation image learned from Section~\ref{sec: lossfieldimaging}, $\boldsymbol{W}_{\mathcal{T}}$ is the weight matrix using the ellipse model, and $N_{\mathcal{T}}=|\mathcal{T}|$ is the size of the dataset $\mathcal{T}$.


The other metric, \textit{running time}, is a measure of the computational efficiency of the proposed \acronym{} algorithm. It has been crucial in time-sensitive applications such as real-time spectrum access and management systems \cite{durgin2022digital}. This metric includes the execution time for loss field learning and shadowing loss prediction. Note that the terms ``learning'' and ``training'', ``prediction'' and ``testing'' are used interchangeably for comparing \acronym{} to the selected approaches in Section~\ref{sec: results}.


We take the following three steps to ensure result comparability. First, all the models are trained and tested on the same partitioned datasets.
Second, the inputs of these ML models are the 2D coordinates of transmitters and receivers to be consistent with \acronym{}. Lastly, all the results are obtained by running the algorithm on the same Linux system with a 16-core Intel Xeon Gold 6130 processor.

\section{Results}\label{sec: results}
Experimental results of the proposed \acronym{} algorithm are given in this section. We first present two loss field image examples which are learned from the datasets in Section~\ref{subsec: datasets}. We then compare \acronym{} with the chosen models via \textit{variance reduction and latency} from Section~\ref{model:evaluation}. The impact of the hyperparameters on accuracy is also discussed. Finally, we present results on the measurement noise variance and the small-scaling fading loss variance. The combination of the two approximates the total noise variance as a lower bound for the fading loss variance.

\subsection{Example loss field images}

This subsection presents two example loss field images using the \textit{log-distance path loss} model in Section~\ref{sec: pathlossfading} and the proposed \acronym{} algorithm in Section~\ref{sec: lossfieldimaging}. They are learned from the \textit{Rooftop} outdoor and indoor training datasets respectively. The rationale behind the \textit{Rooftop} dataset choice is that these receivers, as deployed high above the ground, give better coverage of the campus area. We select both outdoor and indoor datasets to discuss \acronym{}'s practical use in various types of environments. The image boundaries are the same as Fig.~\ref{fig:frs_dataset_transmitter} and \ref{fig:indoor_dataset}.

The statistical analysis follows the next four steps. First, we determine the path loss exponent $n_p$ and the reference loss $P_T-\Pi_0$ in (\ref{model: pathloss}) via linear regression. The reference distance $\Delta_0$ is set to be 1~m across the datasets. The results of the two examples are (1) \textit{Rooftop}: $n_p = 2.73$ and $P_T-\Pi_0 = -1.25$~dB, and (2) indoor: $n_p = 2.26$ and $P_T-\Pi_0 = -37.04$~dB. 

\begin{table}
      {\begin{tabular}{FHAL} 
    \toprule
    Hyperparameter & Description & \textit{Rooftop} & Indoor\\
    \midrule
    $\delta_p$ &Pixel width (m) &  25 & 0.35\\
    \hline
    ${\sigma_{\boldsymbol{x}}^2}/{\sigma_{\boldsymbol{z}}^2}$ & Shadowing Variance Ratio & 0.58 & 0.30\\
    \hline
    $\delta$ & Space Constant & 35 & 2.5\\
    \hline
    $\lambda$ & Excess Length (m) & 105 & 0.18\\
    \hline
    $\alpha$ & Regularization & 0.3 & 41\\
    \bottomrule
  \end{tabular}}
  \caption{Model hyperparameters for \acronym{}.}
  \label{imageexmaple: hyperparameters}
\end{table}

Second, we tune hyperparameters for \acronym{} and interpret their values. The model hyperparameters are selected via 5-fold cross-validation. This procedure is to randomly sample ${1}/{5}$ data out of the training dataset for hyperparameter validation and overfitting prevention. Their descriptions and values are given in Table~\ref{imageexmaple: hyperparameters}. The first hyperparameter, $\delta_p$, denotes the attenuation image resolution and impacts both computation time and prediction accuracy. The second shadowing variance ratio, ${\sigma_{\boldsymbol{x}}^2}/{\sigma_{\boldsymbol{z}}^2}$, represents the contribution of shadowing loss to the total fading loss. In comparison to outdoor environments, indoor surroundings have more multipath components as indoor obstacles like walls that obstruct radio wave propagation are relatively uniformly placed throughout the area. Therefore the indoor dataset shows less variation in shadowing.
The third space constant $\delta$ indicates the obstruction size in the environment \cite{Piyush2009Correlated}. We expect that obstacles will be smaller for the indoor area. In this case, the $\delta$ 
for the \textit{Rooftop} dataset is 35, larger than 2.5 for the indoor dataset. The next hyperparameter $\lambda$ is introduced by the ellipse weight model to select valid pixels for each link. It is determined by the area size and the pixel width. The last hyperparameter $\alpha$ balances the loss field prior and the data from the area of interest. We notice that $\alpha$ 
of the indoor dataset is about 100 times larger than that of the outdoor case. This can be explained by the ${1}/{\sqrt{d_l}}$ weight in \ref{totalfading:weightmatrix}. The path lengths $d_l$ of the indoor measurements are 100 times smaller, which makes $\alpha$ 100 times larger to balance the ${1}/{{d_l}}$ discrepancy in \ref{image_estimator}.


Next, we derive the weight matrix and estimate the loss image via Bayesian linear regression. Fig.~\ref{fig: LossField} demonstrates the two trained loss images and the site maps as a reference. It can be observed that they have spatial loss ranges of -24--24~dB and -1.25--1.00~dB respectively. 
Higher losses can be seen at higher obstructions such as the marked rectangle areas in Fig.~\ref{fig:cbrssdr_LossField} and near cubicle walls in Fig.\ref{fig:indoor_LossField}. 

The red ellipse area of Fig.~\ref{fig:cbrssdr_LossField} highlights a mismatch between the estimated two high-loss regions and one high obstruction of the site map. The loss image estimate is in fact more accurate because the terrain
profile is outdated; a new building recently constructed at the star ($\star$) location was not in the database used to generate the left image in Fig.~\ref{fig:cbrssdr_LossField}. Note that \acronym{} does not use any terrain or building information. Collecting and maintaining the site-specific terrain dataset could be time-consuming and expensive, but \acronym{} can use channel loss measurements for accurate and cost-effective loss field estimation. 


The correlation between obstructions and spatial losses is further proposed for wall imaging \cite{karanam20173d}.
Fig.~\ref{fig:indoor_LossField} presents the loss image of the indoor office and the cubicle locations. It can be observed that desks, computers, and bookcases are generally positioned close to the cubicle walls. Correspondingly the estimated loss image is lower in the middle of each cubicle and higher close to the cubicle walls where these obstructions are more often placed. Similarly, the vertical corridor region at $x\approx3.2$~m experiences lower losses than either side of the corridor. The edges of the loss image are generally close to zero due to the lack of measurements and thus the estimates in that region mostly rely on the field prior. The match between the environment and the loss image validates that the proposed \acronym{} approach has the potential for spatial loss field learning and further shadowing loss prediction. 


The final step is to quantitatively assess the accuracy of the learned loss image via variance reduction. For the \textit{Rooftop} training dataset, the fading loss variance after the \textit{log-distance path loss model} is 58.4 dB$^2$. The MSE by estimating the shadowing loss decreases to 30.7~dB$^2$ which is 47.4\% less than that of the base model. For the indoor training dataset, the fading loss variance reduces from 19.8~dB$^2$ to 10.1~dB$^2$, which corresponds to a 49.3\% reduction. 

\begin{table*}
\centering
\begin{tabular}{cccccccccc}\toprule
\multirow{2}{*}{Receiver} &\multicolumn{4}{c}{Training Time (s)} &\multicolumn{5}{c}{Testing Time (s)} \\\cmidrule(lr){2-5}\cmidrule(lr){6-10}
&Random Forest &SVR &MLP-ANN &CELF &Okumura-Hata &Random Forest &SVR &MLP-ANN &CELF \\\midrule
Rooftop &1.210 &7.411 &26.753 &8.215 &0.001 &0.018 &0.726 &0.011 &0.402 \\
Fixed &1.587 &28.687 &59.671 &15.767 &0.001 &0.027 &2.353 &0.024 &0.642 \\
Mobile &1.605 &4.247 &13.552 &4.837 &0.001 &0.014 &0.311 &0.004 &0.243 \\
Dense &0.700 &7.105 &25.524 &6.036 &0.001 &0.014 &0.755 &0.005 &0.357 \\
DS-SS &0.131 &0.041 &2.402 &0.133 &-- &0.007 &0.006 &0.002 &0.005 \\
\bottomrule
\end{tabular}
\caption{Running time comparison for training and testing among \textit{Okumura-Hata}, ML models and the \acronym{} algorithm.}
\label{tab: latency_runtime}
\end{table*}

\subsection{Accuracy analysis}
\begin{figure}
    \centering
    \includegraphics[width=\columnwidth]{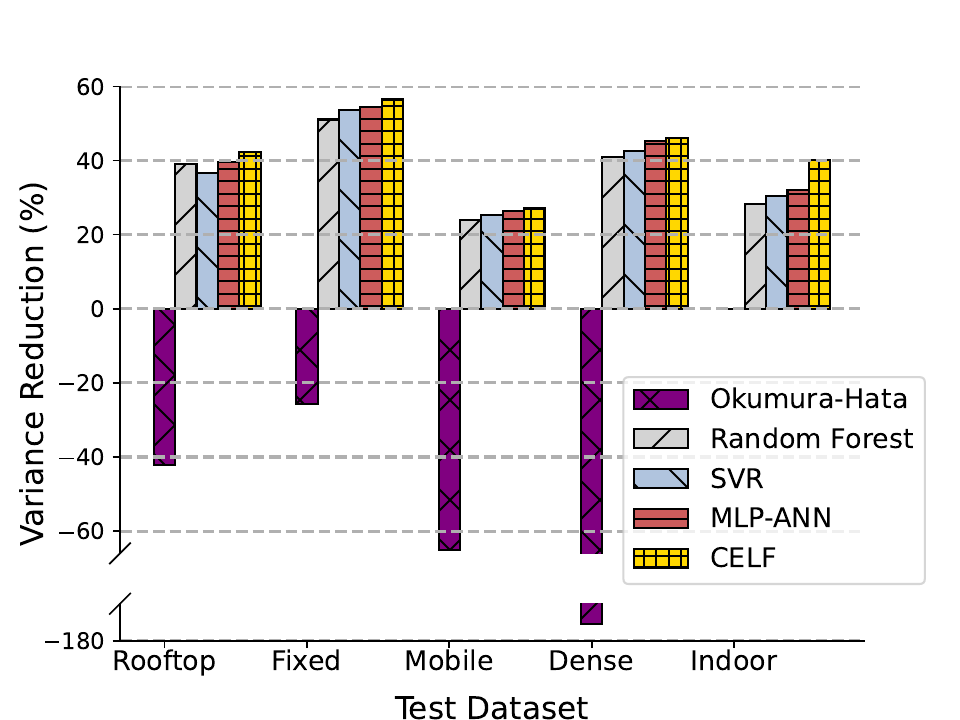}
    \caption{Variance reductions on the outdoor and indoor test datasets via \textit{Okumura-Hata}, three ML methods, and \acronym{}. 
    }
    \label{fig: variancereduction}
\end{figure}


Upon obtaining the loss image, we evaluate \acronym{}'s performance on the test datasets. The first is the accuracy analysis using the variance reduction metric. Fig.~\ref{fig: variancereduction} demonstrates the variance reduction results on the outdoor and indoor test datasets. Note that the \textit{Okumura-Hata} model predicts path loss directly without first-order channel estimation and thus its error variance is computed as the \textit{unbiased path loss variance}. It can be seen from Fig.~\ref{fig: variancereduction} that all methods except the \textit{Okumura-Hata model} can lower the fading loss variance to a certain degree. MLP-ANN gives the largest variance reduction among the three ML-based methods. However, \acronym{} outperforms all the ML models across the test datasets. Take the \textit{Rooftop} dataset for instance. \acronym{} can achieve 42.3\% variance reduction which is higher than MLP-ANN's 39.6\%. To summarize, we are able to show that the \acronym{} algorithm outperforms the three ML methods in terms of variance reduction.

\subsection{Efficiency analysis}
We compare the training and testing efficiency of the methods via running time. Their results are shown in Table ~\ref{tab: latency_runtime}. First, it can be observed that the \textit{Okumura-Hata model} provides the fastest predictions due to its closed-form computation. Second, MLP-ANN, among the remaining methods, is the most efficient for shadowing loss prediction but the most computationally expensive for training. Third, the slowest model for testing is SVR except for the indoor dataset. Last, \acronym{} is approximately 3 times faster than MLP-ANN for image learning. As a result, it can update the model with new measurements or learn the spatial loss of a new environment with much less computational cost. Comparing the prediction time, we can see that \acronym{} is slower than MLP-ANN across all the datasets. This is due to the time-expensive weight matrix computation for each data point. Optimization of the weight model is needed and remains future work.

\subsection{Effect of hyperparameters}
\acronym{}'s hyperparameters play a significant role in its performance. We here present variance reduction as a function of \acronym{}'s three major hyperparameters on the indoor dataset.

\begin{figure*}[t]
  \begin{subfigure}[b]{0.33\textwidth}
    \includegraphics[width=\textwidth]{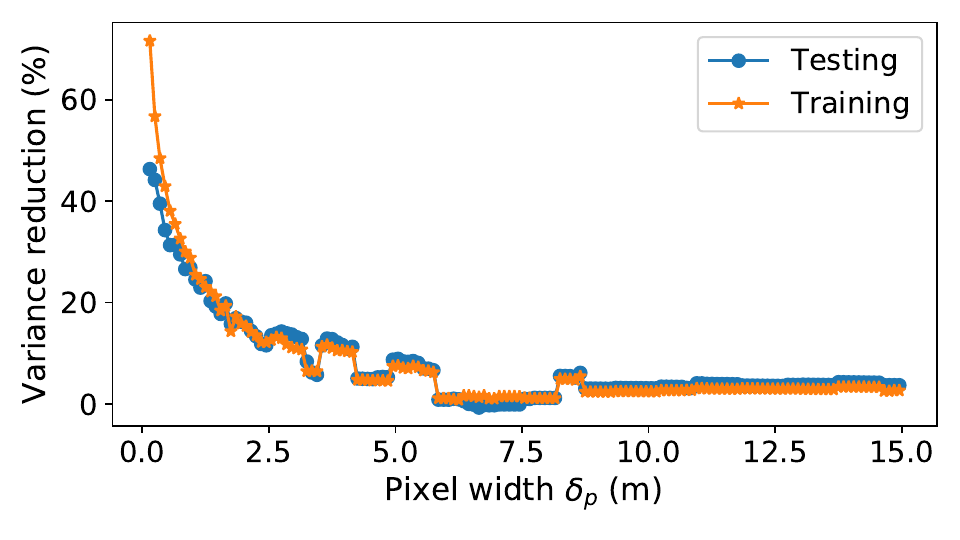}
    \caption{The effect of the pixel width $\delta_p$.}
    \label{fig:PixelWidthEffect}
  \end{subfigure}
  \begin{subfigure}[b]{0.33\textwidth}
    \includegraphics[width=\textwidth]{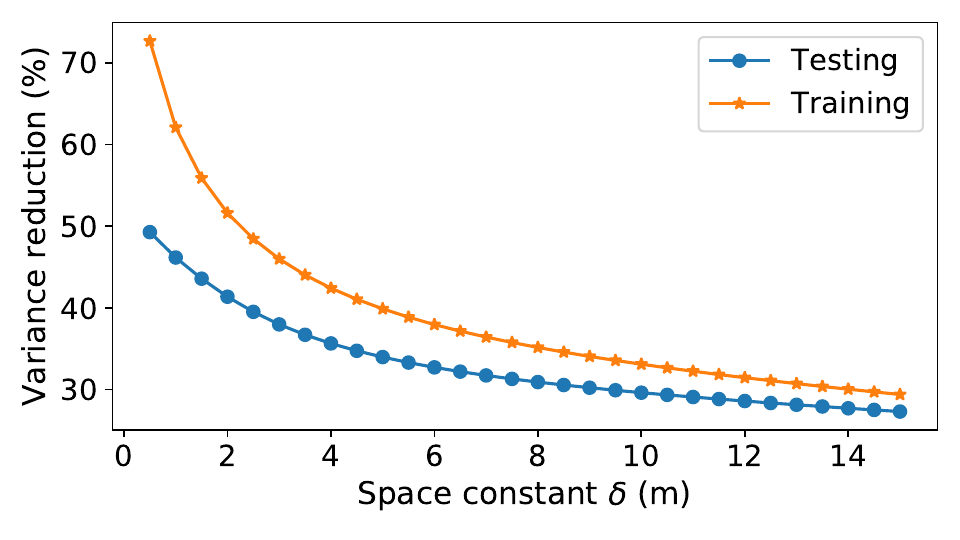}
    \caption{The effect of the space constant $\delta$.}
    \label{fig:SpaceConstantEffect}
  \end{subfigure}
  \begin{subfigure}[b]{0.33\textwidth}
    \includegraphics[width=\textwidth]{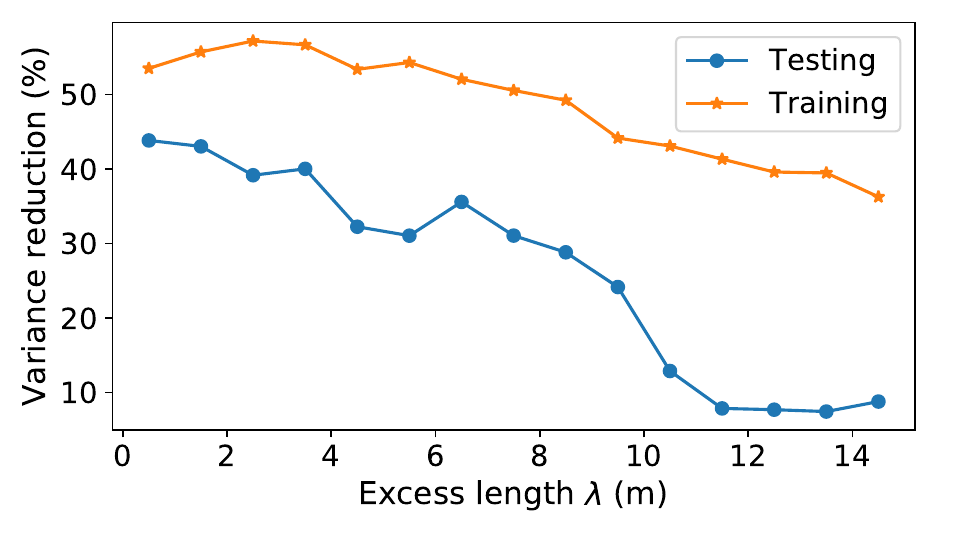}
    \caption{The effect of the excess length $\lambda$.}
    \label{fig:ExcessLengthEffect}
  \end{subfigure}
\caption{Variance reduction vs. \acronym{}'s hyperparameters on the indoor dataset.}
\label{fig: hyperparameter_effects}
\end{figure*}

Fig.~\ref{fig:PixelWidthEffect} shows that the variances for both training and test datasets mostly reduce less as the pixel width $\delta_p$ increases from 0.15m to 15m. Fluctuations occur at near 2.5~m, 3.5~m, 5~m, and 8~m. While the lower the pixel width, the higher the variance reduction, it comes with a training time sacrifice. Fig.~\ref{fig:SpaceConstantEffect} discusses the reduction variation vs. the space constant $\delta$. Reductions for testing and training decrease as $\delta$ increase from 0.5m to 15m. As $\delta$ approximates the obstruction size, unreasonable large space constants give lower variance reduction for training and testing. Fig.~\ref{fig:ExcessLengthEffect} presents the effect of the excess length $\lambda$ on variance reduction. It can be seen that too large of the excess length includes too many pixels for loss field estimation and thus leads to lower variance reduction.


\subsection{Lower variance bound approximation}

We analyze a subset of the outdoor dataset which is collected when the FM transmitter is either stationary or rotating with a radius less than or equal to 1 wavelength ($\lambda_f$). The subset has 14,026 received power observations. Variation in stationary data approximates the measurement noise variance, and the data for link distances changing on the order of the signal wavelength can estimate the small-scale fading loss \cite{goldsmith2005wireless}. Hence the sum of the two gives a sense of the lower bound on the total fading loss variance $\sigma^2_{\boldsymbol{n}}$. Table~\ref{tab: datavariance} illustrates the variance of the two measurement sets. Note that the \textit{Mobile} dataset is not applicable as the receivers are constantly moving. We can learn that for the \textit{Dense} dataset, the variance reduction upper limit is 58.8\% which, based on Fig.~\ref{fig: variancereduction}, is 12.9\% higher than the result of \acronym{}. By comparing Table~\ref{tab: datavariance} and Fig.~\ref{fig: variancereduction}, we can conclude that there is still room to lower the shadowing loss variance, but the proposed method has shown results closer to the limits.

\begin{table}[H]
\centering
\begin{tabular}{lcccc}\toprule
    \multirow{2}{*}{Receiver} &\multicolumn{3}{c}{Data Variance (dB$^2$)} &Reduction \\\cmidrule{2-4}
    &Stationary &Radius$\le1\lambda_f$ &Sum & Percentage\\\midrule
    Rooftop &3.9	&20.3	&24.3	&58.4\% \\
    Fixed &2.9	&11.0	&13.9	&76.9\% \\
    Dense &2.6	&8.2	&10.8	&58.8\% \\
    \bottomrule
\end{tabular}
\caption{Data variance when the portable transmitter is stationary or rotating with a radius $\le 1\lambda_f$.}\label{tab: datavariance}
\end{table}
\section{Related Work}\label{sec:related}

Path loss prediction has an extensive disciplinary history over many decades.  Models used today vary by to what extent they rely on:
\begin{enumerate}
    \item the physical mechanisms of radio propagation, e.g., reflection and diffraction; 
    \item information about the site, e.g., terrain and building geometry data;
    \item curve-fitting to empirical data recorded in past measurements;
    \item fitting or learning using empirical data collected in the area of deployment.
\end{enumerate} 
While some models do not characterize the probability distribution of the channel loss, \textit{statistical models} state a distributional model for the loss variation.


\subsection{Physics-based models} Physics-based models aim to accurately characterize radio wave propagation effects such as reflection, and diffraction. The most fundamental is the \textit{free-space path loss model}\cite{rappaport2010wireless}, but it models only unobstructed channels, and is thus limited to satellite communication and unobstructed microwave relay links. The \textit{two-ray ground reflection model} accounts for both the line-of-sight (LOS) and the ground-reflected paths\cite{phillips2012survey}, and is typically used in flat clutter-free areas like plains\cite{goldsmith2005wireless}. When more multipath must be modeled, \textit{ray tracing} is both the most accurate and complicated model for path loss \cite{yun2015ray}. Ray tracing requires site-specific building databases, i.e., building layout, heights, and dielectric properties, as well as detailed terrain and ground use data, so that each wave path can be traced using geometrical optics \cite{hrovat2013survey}. Its computational complexity and need for high-resolution site-specific data make it impractical for large-scale, real-time applications.

\subsection{General empirical models} General empirical models are based on an analysis of measurements taken from an environment similar in use to the area of interest, e.g., urban or suburban. The \textit{Okumura-Hata model} is based on measurements from Tokyo in the 1960s as formulated by Hata \cite{hata1980empirical}.  It uses curve-fitting to model the effect of signal frequency, antenna heights, path length, and environment type on the channel loss. 
The \textit{COST-231 Hata model} extends the Okumura-Hata model to data from some European cities. 
\cite{seybold2005introduction}. The benefits of statistical models are the simple closed-form formula and no need for data from the site of interest. However, they are restricted to certain frequency ranges, distance ranges, and, most critically, to the environments from which the measurements came \cite{goldsmith2005wireless}. 

\subsection{Hybrid empirical/physical models} The  \textit{Longley-Rice model} or the irregular terrain model (ITM)  combines empirical modeling and physical principles for ground reflection, knife-edge and far-field diffraction, and troposcatter predictions \cite{hufford1982guide}. This model 
considers environmental factors including surface refractivity, ground conductivity, atmospheric parameters, and terrain irregularities for path loss prediction \cite{yun2015ray}. It is in use today by various applications like SAS \cite{Souryal2019Effect}. The terrain-integrated rough Earth model (TIREM)  model \cite{eppink1994tirem} considers a profile of the terrain features and building heights for model improvement \cite{varner2022enhanced}. The last hybrid model is \textit{the ITU-R P.1812 model} is one of the Intl. Telecommunication Union (ITU) terrain models which target path-specific predictions using detailed terrain profiles. It has been widely used for terrestrial wireless  systems \cite{ITU2021model1812}. 


\subsection{Statistical Models} They characterize the statistical distribution of the channel losses, rather than only the average value.  The most common model is the \textit{log-normal shadowing model}, which models shadowing loss as normally distributed in dB \cite{goldsmith2005wireless}. Other models explain the statistical correlation between the shadowing loss on two proximate links \cite{szyszkowicz2010feasibility, lee2018effect, patwari2008nesh}, which become correlated by passing through the same or similar obstructions. \acronym{} models this correlation implicitly via its loss field.
Other distributions for shadowing include the \textit{Nakagami-m} \cite{beaulieu2005efficient} and \textit{Weibull} \cite{sagias2005gaussian} distributions.
We note that the most well-known distributions, Rayleigh and Rician, are models for small-scale fading loss, and are thus not further discussed in this paper.




Another popular class is ML channel models which are designed using general-purpose ML architectures and extensive datasets \cite{seretis2021overview, jo2020path, zhang2019path}. We categorize these models as: (1) SVR, K-Nearest-Neighbors (KNN), and ensemble learning methods such as random forests \cite{zhang2019path}; (2) ANN models including MLP-ANN models \cite{wu2020artificial, ostlin2010macrocell} and radial basis function-ANN models (RBF-ANN) \cite{ojo2021radial}, and (3) more complex DNN models \cite{levie2021radiounet, thrane2020model}. For example, the RadioUNet model in \cite{levie2021radiounet} utilizes large datasets and environmental geometry as input to Unet, a special Convoluted Neural Network (CNN) architecture for path loss modeling.

ML-based methods can provide higher prediction accuracy than domain-specific models at the cost of extensive datasets or detailed environmental information. Additionally, the high complexity of model training and updating will result in significant latency.
The lack of interpretability of ML methods further reduces trust, as RF engineers can find it difficult to diagnose the problem when there is poor model performance. \acronym{} is also a learning-based model which uses site measurements to train. It requires no knowledge about the environment and can be trained with fewer measurements than a general-purpose ML model. Further, \acronym{} explains its estimates via the shadowing field image, which should correlate to the RF obstructions in the area.  

\section{Conclusion}\label{sec:conclusion}

This paper proposes \acronym{} to learn a spatial loss field and predict shadowing loss on new links for channel and transmit power assignment. It formulates total fading loss as a discretized linear model and applies Bayesian linear regression and optimization for the loss image estimation. 

The proposed method has been validated with two evaluation metrics, variance reduction, and running time for training and prediction. It is tested on one outdoor and one indoor real-world dataset. The \textit{Okumura-Hata} model and three ML-based methods, SVR, random forest, and MLP-ANN, are used for performance comparison. Experimental results demonstrate that \acronym{} presents larger variance reductions than all the other methods and can also estimate the loss image more efficiently than the most accurate MLP-ANN model. 

\bibliographystyle{IEEEtran}
\bibliography{main}

\begin{thebibliography}{10}
\providecommand{\url}[1]{#1}
\csname url@samestyle\endcsname
\providecommand{\newblock}{\relax}
\providecommand{\bibinfo}[2]{#2}
\providecommand{\BIBentrySTDinterwordspacing}{\spaceskip=0pt\relax}
\providecommand{\BIBentryALTinterwordstretchfactor}{4}
\providecommand{\BIBentryALTinterwordspacing}{\spaceskip=\fontdimen2\font plus
\BIBentryALTinterwordstretchfactor\fontdimen3\font minus \fontdimen4\font\relax}
\providecommand{\BIBforeignlanguage}[2]{{%
\expandafter\ifx\csname l@#1\endcsname\relax
\typeout{** WARNING: IEEEtran.bst: No hyphenation pattern has been}%
\typeout{** loaded for the language `#1'. Using the pattern for}%
\typeout{** the default language instead.}%
\else
\language=\csname l@#1\endcsname
\fi
#2}}
\providecommand{\BIBdecl}{\relax}
\BIBdecl

\bibitem{ahmad20205g}
W.~S. H. M.~W. Ahmad, N.~A.~M. Radzi, F.~Samidi, A.~Ismail, F.~Abdullah, M.~Z. Jamaludin, and M.~Zakaria, ``5{G} technology: Towards dynamic spectrum sharing using cognitive radio networks,'' \emph{IEEE access}, vol.~8, pp. 14\,460--14\,488, 2020.

\bibitem{bhattarai2016overview}
S.~Bhattarai, J.-M.~J. Park, B.~Gao, K.~Bian, and W.~Lehr, ``An overview of dynamic spectrum sharing: Ongoing initiatives, challenges, and a roadmap for future research,'' \emph{IEEE Transactions on Cognitive Communications and Networking}, vol.~2, no.~2, pp. 110--128, 2016.

\bibitem{sohul2015spectrum}
M.~M. Sohul, M.~Yao, T.~Yang, and J.~H. Reed, ``Spectrum access system for the citizen broadband radio service,'' \emph{IEEE Communications Magazine}, vol.~53, no.~7, pp. 18--25, 2015.

\bibitem{kidd2018national}
\BIBentryALTinterwordspacing
T.~Kidd, ``National radio quiet and dynamic zones,'' 2018. [Online]. Available: \url{https://www.doncio.navy.mil/chips/ArticleDetails.aspx?ID=10299}
\BIBentrySTDinterwordspacing

\bibitem{yun2015ray}
Z.~Yun and M.~F. Iskander, ``Ray tracing for radio propagation modeling: Principles and applications,'' \emph{IEEE access}, vol.~3, pp. 1089--1100, 2015.

\bibitem{eppink1994tirem}
D.~Eppink and W.~Kuebler, ``{TIREM/SEM} handbook,'' \emph{Defense Technical Information Center}, 1994.

\bibitem{hata1980empirical}
M.~Hata, ``Empirical formula for propagation loss in land mobile radio services,'' \emph{IEEE transactions on Vehicular Technology}, vol.~29, no.~3, pp. 317--325, 1980.

\bibitem{rappaport2010wireless}
T.~S. Rappaport, \emph{Wireless communications: Principles and practice, 2/E}.\hskip 1em plus 0.5em minus 0.4em\relax Pearson Education India, 2010.

\bibitem{goldsmith2005wireless}
A.~Goldsmith, \emph{Wireless communications}.\hskip 1em plus 0.5em minus 0.4em\relax Cambridge university press, 2005.

\bibitem{wilson2010radio}
J.~Wilson and N.~Patwari, ``Radio tomographic imaging with wireless networks,'' \emph{IEEE Transactions on Mobile Computing}, vol.~9, no.~5, pp. 621--632, 2010.

\bibitem{bettstetter2003connectivity}
C.~Bettstetter and C.~Hartmann, ``Connectivity of wireless multihop networks in a shadow fading environment,'' in \emph{Proceedings of the 6th ACM international workshop on modeling analysis and simulation of wireless and mobile systems}, 2003, pp. 28--32.

\bibitem{hekmat2006connectivity}
R.~Hekmat and P.~Van~Mieghem, ``Connectivity in wireless ad-hoc networks with a log-normal radio model,'' \emph{Mobile networks and applications}, vol.~11, pp. 351--360, 2006.

\bibitem{chen2011implications}
Y.~Chen and A.~Terzis, ``On the implications of the log-normal path loss model: an efficient method to deploy and move sensor motes,'' in \emph{Proceedings of the 9th ACM conference on embedded networked sensor systems}, 2011, pp. 26--39.

\bibitem{gudmundson1991correlation}
M.~Gudmundson, ``Correlation model for shadow fading in mobile radio systems,'' \emph{Electronics letters}, vol.~23, no.~27, pp. 2145--2146, 1991.

\bibitem{Piyush2009Correlated}
P.~Agrawal and N.~Patwari, ``Correlated link shadow fading in multi-hop wireless networks,'' \emph{IEEE Transactions on Wireless Communications}, vol.~8, no.~8, pp. 4024--4036, 2009.

\bibitem{lee2018effect}
J.~Lee and F.~Baccelli, ``On the effect of shadowing correlation on wireless network performance,'' in \emph{IEEE INFOCOM 2018-IEEE Conference on Computer Communications}.\hskip 1em plus 0.5em minus 0.4em\relax IEEE, 2018, pp. 1601--1609.

\bibitem{patwari2008effects}
N.~Patwari and P.~Agrawal, ``Effects of correlated shadowing: Connectivity, localization, and {RF} tomography,'' in \emph{2008 International Conference on Information Processing in Sensor Networks ({IPSN} 2008)}.\hskip 1em plus 0.5em minus 0.4em\relax IEEE, 2008, pp. 82--93.

\bibitem{tse2005fundamentals}
D.~Tse and P.~Viswanath, \emph{Fundamentals of wireless communication}.\hskip 1em plus 0.5em minus 0.4em\relax Cambridge university press, 2005.

\bibitem{golub2013matrix}
G.~H. Golub and C.~F. Van~Loan, \emph{Matrix computations}.\hskip 1em plus 0.5em minus 0.4em\relax JHU press, 2013.

\bibitem{garnett2023bayesian}
R.~Garnett, \emph{Bayesian optimization}.\hskip 1em plus 0.5em minus 0.4em\relax Cambridge University Press, 2023.

\bibitem{RFSData2022gitlab}
F.~Mitchell, ``The frs/gmrs outdoor received power dataset,'' \url{https://zenodo.org/record/7259895\#.ZDbPiezML0p}, 2022.

\bibitem{Breenwintech20}
J.~Breen, A.~Buffmire, J.~Duerig, K.~Dutt, E.~Eide, M.~Hibler, D.~Johnson, S.~K. Kasera, E.~Lewis, D.~Maas, A.~Orange, N.~Patwari, D.~Reading, R.~Ricci, D.~Schurig, L.~B. Stoller, J.~Van~der Merwe, K.~Webb, and G.~Wong, ``{POWDER}: Platform for open wireless data-driven experimental research,'' in \emph{Proceedings of the 14th International Workshop on Wireless Network Testbeds, Experimental Evaluation and Characterization (WiNTECH)}, Sep. 2020.

\bibitem{indoor2007patwari}
N.~Patwari, ``Indoor channel impulse response dataset,'' \url{https://crawdad.org/utah/CIR/20070910/matlab/}, 2022.

\bibitem{patwari2003relative}
N.~Patwari, A.~O. Hero, M.~Perkins, N.~S. Correal, and R.~J. O'dea, ``Relative location estimation in wireless sensor networks,'' \emph{IEEE Transactions on signal processing}, vol.~51, no.~8, pp. 2137--2148, 2003.

\bibitem{levie2021radiounet}
R.~Levie, {\c{C}}.~Yapar, G.~Kutyniok, and G.~Caire, ``Radiounet: Fast radio map estimation with convolutional neural networks,'' \emph{IEEE Transactions on Wireless Communications}, vol.~20, no.~6, pp. 4001--4015, 2021.

\bibitem{marey2022pl}
A.~Marey, M.~Bal, H.~F. Ates, and B.~K. Gunturk, ``{PL-GAN}: Path loss prediction using generative adversarial networks,'' \emph{IEEE Access}, vol.~10, pp. 90\,474--90\,480, 2022.

\bibitem{wen2019path}
J.~Wen, Y.~Zhang, G.~Yang, Z.~He, and W.~Zhang, ``Path loss prediction based on machine learning methods for aircraft cabin environments,'' \emph{Ieee Access}, vol.~7, pp. 159\,251--159\,261, 2019.

\bibitem{zhang2019path}
Y.~Zhang, J.~Wen, G.~Yang, Z.~He, and J.~Wang, ``Path loss prediction based on machine learning: Principle, method, and data expansion,'' \emph{Applied Sciences}, vol.~9, no.~9, p. 1908, 2019.

\bibitem{wu2020artificial}
L.~Wu, D.~He, B.~Ai, J.~Wang, H.~Qi, K.~Guan, and Z.~Zhong, ``Artificial neural network based path loss prediction for wireless communication network,'' \emph{IEEE access}, vol.~8, pp. 199\,523--199\,538, 2020.

\bibitem{drocella20163}
E.~F. Drocella, J.~Richards, R.~Sole, F.~Najmy, A.~Lundy, and P.~McKenna, ``3.5 {GH}z exclusion zone analyses and methodology,'' National Telecommunications and Information, Tech. Rep., 2016.

\bibitem{oroza2017machine}
C.~A. Oroza, Z.~Zhang, T.~Watteyne, and S.~D. Glaser, ``A machine-learning-based connectivity model for complex terrain large-scale low-power wireless deployments,'' \emph{IEEE Transactions on Cognitive Communications and Networking}, vol.~3, no.~4, pp. 576--584, 2017.

\bibitem{moraitis2020machine}
N.~Moraitis, L.~Tsipi, and D.~Vouyioukas, ``Machine learning-based methods for path loss prediction in urban environment for {LTE} networks,'' in \emph{2020 16th international conference on wireless and mobile computing, networking and communications (WiMob)}.\hskip 1em plus 0.5em minus 0.4em\relax IEEE, 2020, pp. 1--6.

\bibitem{durgin2022digital}
G.~D. Durgin, M.~A. Varner, N.~Patwari, S.~K. Kasera, and J.~Van~der Merwe, ``Digital spectrum twinning for next-generation spectrum management and metering,'' in \emph{2022 IEEE 2nd International Conference on Digital Twins and Parallel Intelligence (DTPI)}.\hskip 1em plus 0.5em minus 0.4em\relax IEEE, 2022, pp. 1--6.

\bibitem{karanam20173d}
C.~R. Karanam and Y.~Mostofi, ``3{D} through-wall imaging with unmanned aerial vehicles using {W}i{F}i,'' in \emph{Proceedings of the 16th ACM/IEEE International Conference on Information Processing in Sensor Networks}, 2017, pp. 131--142.

\bibitem{phillips2012survey}
C.~Phillips, D.~Sicker, and D.~Grunwald, ``A survey of wireless path loss prediction and coverage mapping methods,'' \emph{IEEE Communications Surveys \& Tutorials}, vol.~15, no.~1, pp. 255--270, 2012.

\bibitem{hrovat2013survey}
A.~Hrovat, G.~Kandus, and T.~Javornik, ``A survey of radio propagation modeling for tunnels,'' \emph{IEEE Communications Surveys \& Tutorials}, vol.~16, no.~2, pp. 658--669, 2013.

\bibitem{seybold2005introduction}
J.~S. Seybold, \emph{Introduction to {RF} propagation}.\hskip 1em plus 0.5em minus 0.4em\relax John Wiley \& Sons, 2005.

\bibitem{hufford1982guide}
G.~A. Hufford, A.~G. Longley, W.~A. Kissick \emph{et~al.}, \emph{A guide to the use of the ITS irregular terrain model in the area prediction mode}.\hskip 1em plus 0.5em minus 0.4em\relax US Department of Commerce, National Telecommunications and Information Administration, 1982.

\bibitem{Souryal2019Effect}
M.~R. Souryal and T.~T. Nguyen, ``Effect of federal incumbent activity on {CBRS} commercial service,'' in \emph{2019 IEEE International Symposium on Dynamic Spectrum Access Networks (DySPAN)}, 2019, pp. 1--5.

\bibitem{varner2022enhanced}
M.~A. Varner, F.~Mitchell, J.~Wang, K.~Webb, and G.~D. Durgin, ``Enhanced {RF} modeling accuracy using simple minimum mean-squared error correction factors,'' in \emph{2022 IEEE 2nd International Conference on Digital Twins and Parallel Intelligence (DTPI)}.\hskip 1em plus 0.5em minus 0.4em\relax IEEE, 2022, pp. 1--5.

\bibitem{ITU2021model1812}
P.~Series, \emph{Recommendation ITU-R P.1812-6: A path-specific propagation prediction method for point-to-area terrestrial services in the frequency range 30 MHz to 6000 MHz}.\hskip 1em plus 0.5em minus 0.4em\relax International Telecommunication Union, 2021.

\bibitem{szyszkowicz2010feasibility}
S.~S. Szyszkowicz, H.~Yanikomeroglu, and J.~S. Thompson, ``On the feasibility of wireless shadowing correlation models,'' \emph{IEEE Transactions on Vehicular Technology}, vol.~59, no.~9, pp. 4222--4236, 2010.

\bibitem{patwari2008nesh}
N.~Patwari and P.~Agrawal, ``Nesh: A joint shadowing model for links in a multi-hop network,'' in \emph{2008 IEEE International Conference on Acoustics, Speech and Signal Processing}.\hskip 1em plus 0.5em minus 0.4em\relax IEEE, 2008, pp. 2873--2876.

\bibitem{beaulieu2005efficient}
N.~C. Beaulieu and C.~Cheng, ``Efficient {N}akagami-m fading channel simulation,'' \emph{IEEE Transactions on Vehicular Technology}, vol.~54, no.~2, pp. 413--424, 2005.

\bibitem{sagias2005gaussian}
N.~C. Sagias and G.~K. Karagiannidis, ``Gaussian class multivariate weibull distributions: theory and applications in fading channels,'' \emph{IEEE Transactions on Information Theory}, vol.~51, no.~10, pp. 3608--3619, 2005.

\bibitem{seretis2021overview}
A.~Seretis and C.~D. Sarris, ``An overview of machine learning techniques for radiowave propagation modeling,'' \emph{IEEE Transactions on Antennas and Propagation}, vol.~70, no.~6, pp. 3970--3985, 2021.

\bibitem{jo2020path}
H.-S. Jo, C.~Park, E.~Lee, H.~K. Choi, and J.~Park, ``Path loss prediction based on machine learning techniques: Principal component analysis, artificial neural network, and {G}aussian process,'' \emph{Sensors}, vol.~20, no.~7, p. 1927, 2020.

\bibitem{ostlin2010macrocell}
E.~Ostlin, H.-J. Zepernick, and H.~Suzuki, ``Macrocell path-loss prediction using artificial neural networks,'' \emph{IEEE Transactions on Vehicular Technology}, vol.~59, no.~6, pp. 2735--2747, 2010.

\bibitem{ojo2021radial}
S.~Ojo, A.~Imoize, and D.~Alienyi, ``Radial basis function neural network path loss prediction model for {LTE} networks in multitransmitter signal propagation environments,'' \emph{International Journal of Communication Systems}, vol.~34, no.~3, 2021.

\bibitem{thrane2020model}
J.~Thrane, D.~Zibar, and H.~L. Christiansen, ``Model-aided deep learning method for path loss prediction in mobile communication systems at 2.6 ghz,'' \emph{Ieee Access}, vol.~8, pp. 7925--7936, 2020.

\end{thebibliography}

\end{document}